\documentclass[traditabstract]{aa}

\usepackage{graphicx}
\usepackage[varg]{txfonts}
\usepackage{xcolor}
\usepackage{hyperref}
\usepackage{graphicx}
\usepackage{subcaption}
\usepackage{footnote}
\usepackage{subcaption}

\newcommand{\dvn}{$\Delta v_{90}$}
\newcommand{\tot}{$_{\rm tot}$}
\newcommand{\nucl}{$_{\rm nucl}$}
\newcommand{\kms}{km s$^{-1}$}

\renewcommand{\thefootnote}{\ifcase\value{footnote}
  \or $\blacklozenge$    % 1
  \or $\aleph$           % 2
  \or $\dagger$          % 3
  \or $\ddagger$         % 4
  \or $\mathparagraph$   % 5
  \or $\otimes$          % 6
  \or $\circledast$      % 7
  \or $\bumpeq$          % 8
  \or $\Join$            % 9
  \or $\sharp$           % 10
  \or $\flat$            % 11
  \or $\natural$         % 12
  \else \arabic{footnote}\fi}

\begin{document} 

\title{Chemodynamical properties of gas-rich galaxies: a comparison of observations and simulations}
%\subtitle{}

\author{ Anna~Velichko\inst{\ref{inst1},\ref{inst2}}
\and Yves~Revaz\inst{\ref{inst3}}
\and Annalisa~De~Cia\inst{\ref{inst4}}
\and C\'{e}dric~Ledoux\inst{\ref{inst5}}
\and Jens-Kristian Krogager\inst{\ref{inst6},\ref{inst7}}
\and C\'{e}line P\'{e}roux \inst{\ref{inst4}}
\and Benedetta Casavecchia \inst{\ref{inst8}}
}

\institute{ 
Institute of Astronomy, Kharkiv National University, Svobody sq. 4, Kharkiv, 61022, Ukraine\label{inst1}
\email{velichko.anna.b@gmail.com}\label{email} 
\and 
Department of Astronomy, University of Geneva, Chemin Pegasi 51, 1290 Versoix, Switzerland\label{inst2}
\and
Institute of Physics, Laboratory of Astrophysics, \'Ecole Polytechnique F\'ed\'erale de Lausanne (EPFL), 1290 Sauverny, Switzerland \label{inst3}
\and
European Southern Observatory, Karl-Schwarzschild Str. 2, 85748 Garching bei M\"unchen, Germany\label{inst4}
\and
European Southern Observatory, Alonso de C\'ordova 3107, Casilla 19001, Vitacura, Santiago, Chile\label{inst5}
\and
French-Chilean Laboratory for Astronomy, IRL 3386, CNRS and U. de Chile, Casilla 36-D, Santiago, Chile\label{inst6}
\and
Centre de Recherche Astrophysique de Lyon, Univ. Claude Bernard Lyon 1, 9 Av. Charles Andre, 69230 St Genis Laval, France\label{inst7}
\and
Max-Planck-Institut f\"ur Astrophysik, Karl-Schwarzschild-Str. 1, 85748 Garching b. M\"unchen, Germany\label{inst8}
}

\date{Received date /
Accepted date }

\abstract
{
We perform a comprehensive analysis of the chemical and dynamical properties of quasar-DLA galaxies 
and compare these to the \textsc{GEAR}
chemodynamical simulations. Specifically, we aim to constrain the behavior of $\alpha$-element enhancements with metallicity, the dependence of [$\alpha$/Fe] on the specific star formation rate (sSFR), as well as the absorption-line velocity widths (\dvn) vs. stellar mass, \dvn\, vs. metallicity, and mass–metallicity relations. For the comparison, we select five galaxies simulated with the chemodynamical Tree/SPH code GEAR with stellar masses in the range $6.1\leq$ log $M_\star/M_\odot\leq10.8$ and at six different redshifts between 0.33 and 4.12. We find that the abundance ratios [$\alpha$/Fe] and [M/H] observed in the interstellar medium (ISM) of DLA galaxies overlap with the abundance trends in gas of the simulated galaxies. Our findings corroborate a picture in which DLAs with \dvn\, below and above 100 \kms\, trace galaxies with masses in the ranges $6<\,$log $M_\star<8$ and $8<\,$log $M_\star<11$, respectively. We suggest that observations should be used with caution when constraining the theoretical [$\alpha$/Fe] vs. sSFR relations because of systematics (if abundances are obtained from emission lines) or differences in the gas properties as probed by a DLA and its counterpart. So far, only the observations in absorption of inner gas of the LMC and SMC are in agreement with the simulated data. We confirm that DLAs detected at large impact parameters most likely probe the gas of satellite or other halo galaxies which are adjacent to the central galaxy.
We further find that the velocity widths vs. stellar mass, and mass–metallicity relations agree well with observations, while GEAR should be calibrated more carefully to reproduce the  \dvn\, vs. metallicity relation. To place our results in context, we additionally incorporate chemodynamical properties of a few selected model galaxies obtained from other simulations.
}


   \maketitle

\section{Introduction}

The chemical evolution of galaxies is a complex problem that involves many astrophysical aspects such as metal abundances in stars and gas, gas content and gas flows, amount of dust, stellar masses,  star formation rate (SFR), stellar feedback,  interaction with other galaxies, etc. \citep[][]{Tinsley1980, Matteucci2012}. The study of metal enrichment of gas and stars in galaxies, together with information about their stellar masses and SFRs, provides clues to understanding internal processes that govern the chemical evolution of the galaxies. Indeed, 
[$\alpha$/Fe] vs. [M/H] are important diagnostics for the chemical evolution of galaxies
\citep[][]{Kobayashi2020}. Fe and $\alpha$-elements are formed via different channels with different time scales. Core-collapse supernovae (CCSNe) are the main source producing $\alpha$-elements at an early stage of galaxy evolution, which results in enhanced [$\alpha$/Fe]. Type Ia supernovae (SNe Ia), dominating the iron enrichment, begin to contribute with a time delay of $\sim1$ Gyr, and the [$\alpha$/Fe] ratio begin to decrease \citep[e.g.,][]{Worthey1992, McWilliam1997, Matteucci2012}. The turning point in the [$\alpha$/Fe] vs. [M/H] plot that separates these two regimes is called the ``high-$\alpha$ knee''. Theoretically, the position of the knee is sensitive to the early star formation history (SFH), and it should move toward higher metallicity with increasing SFR \citep[][]{Tinsley1979, McWilliam1997}.

The processes taking place in the Milky Way (MW) and nearby galaxies have been studied in details from spectroscopic observations of individual stars \citep[e.g.,][]{McWilliam1997, Tolstoy2009, deBoer2014, Hasselquist2021}. It has been shown that nearby dwarf galaxies have the high-$\alpha$ knees at different metallicities, and this depends not only on the mass of the galaxies, but also on SNe feedback, galactic winds, interaction with other galaxies, which ultimately influences the SFH \citep[][]{Tolstoy2009}.

Unfortunately, in distant galaxies it is normally not possible to obtain information about individual stars\footnote{there are some exceptions: lensing, SNe, GRBs, etc.}. Revealing the chemical properties of distant galaxies can be done from studies of the integrated stellar population or the interstellar medium (ISM).

A unique way to study the chemical properties of the ISM in distant galaxies is from observations of damped Lyman-$\alpha$ \citep[DLA,][]{Wolfe2005} absorbers which are defined as systems with high column densities of neutral hydrogen ($N{\rm(H I)}\geq 2 \times 10^{20}$ cm$^{-2}$) that appear in absorption of quasi-stellar objects (QSOs) spectra. DLAs are fundamental for the study of galaxies, because they bear a bulk of neutral hydrogen in the Universe \citep[e.g.,][]{Wolfe1995} as well as metals at high redshifts \citep[][]{Peroux2020}.
DLAs are associated with galaxies with a wide mass range and including low-mass galaxies \citep[with stellar masses likely down to $10^6 M_\odot$][]{Christensen2014, Krogager2017, Moller2020}, which in number are the majority \citep[][]{Fontana2006}.
Recently, we used gas observed in absorption in QSO-DLA sources to study the chemical evolution of distant galaxies \citep[][]{Velichko2024}. We analyzed [$\alpha$/Fe] and [M/H] in the neutral ISM/CGM of 24 QSO-DLAs at redshifts in the range $1.6<z<3.4$.

The weakspot of studying DLAs is the difficulty in measuring parameters such as luminosity, stellar mass, and SFR of the corresponding galaxies. It is becoming more evident that correlations between these fundamental parameters and metallicity are crucial to understanding the nature of galaxies at different redshifts \citep[e.g.,][]{Tremonti2004, Savaglio2005, Calura2009, Henry2021, Kashino2022, Scholte2024, Chemerynska2024, Curti2024}. To measure them, observations in emission of the DLA counterparts are needed \citep[][]{Fynbo2008}. Many previous attempts have been made to identify the galaxies responsible for the DLA absorption lines \citep[][]{Djorgovski1996, Moller2002, Ma2015, Kulkarni2022}, but only very few have been detected in emission \citep[][]{Krogager2017}.

Alternatively, the properties of DLAs can be constrained by comparing their measured characteristics with those obtained from modeling. %Several approaches to modeling are used in the literature. %each having its own \textbf{scope of applicability.
Analytical models are used in testing the origin of DLAs \citep[][]{Prochaska1997, Jedamzik1998, Maller2001}; they are successful in reproducing general observed characteristics of DLAs such as observed metallicity distributions \citep[][]{Fynbo2008, Krogager2020}, distribution of cold and warm neutral gas \citep[][]{Krogager2020}.
A limitation is that these models depend on the average scaling ratios of galaxies and cannot capture processes related to galactic substructure or gas structures in the intergalactic medium \citep[][]{Rhodin2019}.

Cosmological hydrodynamical simulations seem to be more capacious and efficient in constraining galaxy properties and relations obtained from observations \citep[e.g., ][]{Nagamine2007, Pontzen2008, Fumagalli2011, Cen2012, Rahmati2013, Hummels2013, Faucher-Gigure2015, Liang2016, Revaz2018, Roca-Fabrega2021}.  The modern cosmological hydrodynamical simulations have been successfully used to study properties of the ISM \citep[e.g., ][]{Altay2013, Rhodin2019, Hassan2020, Garratt-Smithson2021, Klimenko2023} and CGM \citep[][]{Oppenheimer2012, Stinson2012, Shen2012, Shen2013, Oppenheimer2016} in distant galaxies.
It is challenging to reproduce all the diversity of observed galaxy properties and their evolution with redshift because of the complexity of mechanisms governing the formation and evolution of galaxies. 
Hydrodynamical models must govern interactions between multiple constituents, such as gas inflow, gas cooling, UV-background heating, hydrogen self-shielding, star formation, and associated feedback, which creates gas outflows.

In this work, we perform a comprehensive analysis of the chemical and dynamical properties of DLA galaxies, studied in our previous work \citep[][]{Velichko2024}, as well as found in the literature \citep[][]{Weng2023, Berg2023, Kulkarni2022, Moller2020, Christensen2014, Rahmati2014, Krogager2013, Fynbo2013, Fynbo2011}, by comparing them with simulations developed by \citet{Revaz2012, Revaz2016, Revaz2018} using the chemodynamical Tree/SPH code GEAR. We constrain the observed galaxy properties within the stellar mass range $6.1\leq{\rm log }(M_\star/M_\odot)\leq10.8$ and at redshifts $0.33\leq z\leq4.12$.

The paper is organized as follows. In Sect. \ref{sec:data} we describe the observed and simulated data that we use to compare. The methods we apply to extract physical properties and make mock observations from the simulations are described in Sect. \ref{sec:methodology}. In Sect. \ref{sec:results} we compare observations and simulations of some scaling relations that are relevant for the study of galaxy evolution. The masses of the 24 DLAs analyzed in our previous study were estimated in Sect. \ref{sec:fe_h-mg_fe_discussion} by comparing their observed [$\alpha$/Fe]-metallicity distribution with the corresponding trends predicted by the GEAR simulations.
Finally, the main conclusions are reported in Sect. \ref{sec:conclusions}.

\section{Data}
\label{sec:data}

\subsection{Observed [$\alpha$/Fe] vs. [M/H] in the ISM of DLAs}
\label{sec:data_alpha_Fe-M_H}

In our previous work \citep[][]{Velichko2024} we have analyzed abundance patterns of the neutral ISM in a sample of 110 QSO-DLA sources. From there, we selected 24 objects we called as the ``golden'' sample, for which column densities of Ti and at least one more $\alpha$-element were measured. This requirement was needed to apply the method developed by \citet{DeCia2024} to calculate with the greatest reliability the total (gas + dust) metallicity [M/H]\tot\, as well as relative abundances of chemical elements (O, Mg, Si, S, Ti, Cr, Fe, Ni, Zn, P, and Mn) with respect to iron corrected for the depletion by dust, [X/Fe]\nucl. In particular, the measurement of Ti specifically allows the characterisation of the dust-corrected $\alpha$-elements abundances, and thus the derivation of [$\alpha$/Fe]\nucl. In Sect. \ref{sec:fe_h-mg_fe_discussion} we compare the observed values of [$\alpha$/Fe]\nucl\, vs. [M/H]\tot\, with those obtained from the GEAR simulations.

\subsection{Kinematics as a proxy for the dynamical mass}
\label{sec:vel-met_rel}

The masses of the DLA galaxies we studied in our previous work \citep[][]{Velichko2024} are mostly unknown, with two exceptions \citep[Q2243-605, Q2206-1985, which are the most metal-rich and probably most massive galaxies measured by][]{Moller2020}, because it is very difficult to detect their counterparts in emission. Instead, as a mass indicator, we used velocity widths \dvn\, measured from low-ionization line profiles provided by \citet{Ledoux2006} \citep[see Table 1 from][]{Velichko2024}. The argument for using \dvn\, of DLAs as a proxy for the gravitational masses of their hosts is the existence of the velocity width--metallicity correlation obtained by \citet{Ledoux2006} which is the consequence of an underlying mass--metallicity correlation.
This was validated by \citet{Christensen2014} who measured the masses of a dozen DLA counterparts.
The combined mass--velocity width correlation was also confirmed by \citet{Haehnelt1998} from simulations. According to the virial theorem, line-of-sight galaxy velocities provide a measure of the depth of the potential well \citep[e.g.,][]{Saro2013}. All the correlations have a large scatter naturally caused by a variety of impact parameters, inclination angles, presence of infall/outflow of gas, merging galaxy sub-clumps, dependence on redshift, etc. probed by the observations. Considering all these factors, \dvn\, is only a rough proxy of the galaxy mass. 

\subsection{Simulations}
\label{sec:simulated_data}

\subsubsection{GEAR}

To constrain the properties of the DLA galaxies studied in \citet{Velichko2024}, we use model galaxies simulated with the chemodynamical Tree/SPH code GEAR developed by \citet{Revaz2012} and \citet{Revaz2016} which takes into account the complex treatment of baryon physics.
In a nutshell, the gas cooling is computed using the \textsc{GRACKLE} library \citep{Smith2017}
that contains primordial gas and metal-lines cooling as well as H$_2$. 
UV-background radiation \citep{Haardt2012} and hydrogen self-shielding are both included.
Star formation is modeled using a standard stochastic prescription \citep{Katz1992,Katz1996} 
that reproduces the Schmidt law. 
It is supplemented by a Jeans pressure floor adding a nonthermal term in the equation of state.
The stellar feedback includes both Type Ia and II SNe with yields taken from \citet{Tsujimoto1995}.
Both Ia and II SNe explode as individual stars. For type II, the explosion time is 
precisely the end of the star lifetime.
For SNIa, we use the rates from \citet{Kobayashi2000} where explosions start 
after $300\,\rm{Myr}$ at the earliest, following the star formation with the corresponding IMF.
To prevent an instantaneous radiation of the injected energy,
a delayed cooling method is used. It consists in 
disabling gas cooling for a short period of time \citep{Stinson2006}, here taken as $5\,\rm{Myr}$,
after the explosion event.

The GEAR simulations have been tested in the works of \citet{Revaz2018} and \citet{Roca-Fabrega2021}. By comparing with observations of the Local Group dwarf galaxies, \citet{Revaz2018} have shown that the scaling relations between integrated properties of the GEAR simulated galaxies, such as their total V-band luminosity, stellar line-of-sight velocity dispersion, stellar mean metallicity, gas mass content and half-light radius, are well reproduced over several orders of magnitude. Moreover, a good match between six selected Local Group dwarf galaxies and the corresponding GEAR models in detailed stellar properties, such as [Mg/Fe] vs. [Fe/H] (where Mg is used as a proxy of $\alpha$-elements), radial distribution of the stellar line-of-sight velocity dispersion, and the stellar metallicity distribution proves the validity of the GEAR simulations, at least in the low-mass range of $10^6-10^9 M_\odot$ \citep[][]{Revaz2018}. The stellar [Mg/Fe] vs. [Fe/H] relation traces the SFH in galaxies, and the fact that it is well reproduced for the real dwarf galaxies \citep[see Fig. 10 in][]{Revaz2018} shows that the evolution of the stellar population and chemical enrichment implemented in GEAR match well with observations of dwarf galaxies. The MW-mass GEAR galaxy has also been analyzed and compared with other simulations within the Assembling Galaxies Of Resolved Anatomy (AGORA) project \citep[][]{Kim2016, Roca-Fabrega2021} where its ability to simulate MW-like galaxies is demonstrated.

Among 27 GEAR zoom-in dwarf galaxies presented in \citet{Revaz2018}, we select four galaxies with stellar masses in the range of $6.1 \leq$ log $M_\star \leq 8.7$ (at $z=0$): h074, h050, h076, h026 (from least to most massive), in such a way as to cover the low-mass end of the DLA sample described in Sect. \ref{sec:data_alpha_Fe-M_H}. To expand the range under study toward higher masses, we select a MW-like galaxy h000 of log $M_\star = 10.8$ (at $z=0$) from \citet{Roca-Fabrega2021}, also simulated with the GEAR code, but with a slightly stronger feedback than in the dwarf galaxies mentioned above. To trace the galaxy properties over their lifetimes, we analyze six snapshots corresponding to $z =$ 4.12, 3.49, 2.55, 2.01, 1.15 and 0.33. Each snapshot contains information on the physical, chemical, and kinematic properties of stellar and gas particles.
In this work, we focus on the properties of the gas, which is modeled in the simulations as gas ``particles'' with initial masses of a few$\times10^4M_\odot$ \citep[][]{Revaz2018, Roca-Fabrega2021}.
The data also include information on abundances of some chemical elements containing in gas (Fe, Mg, O, S, Zn, Sr, Y, Ba, Eu) calculated with the use of theoretical yields from \citet{Tsujimoto1995, Kobayashi2000}.
Further details on the numerical characteristics of the GEAR models can be found in Appendix \ref{ap:GEAR}.

\subsubsection{Other simulations}

This paper focuses on the comparison of observed DLA properties with GEAR simulated galaxies. Like every simulation, GEAR has its own assumptions and limitations. We include comparison with other simulations to test other physical implementations. % extend beyond the potential limitations of GEAR. 
We found in the literature a few galaxy characteristics, suitable for comparison, obtained by other authors from AURIGA \citep[][]{vandeVoort2019}, IllustrisTNG \citep[][]{Nelson2020, Torrey2019}, performed using the \texttt{AREPO} code, and EAGLE \citep[][]{Matthee2018} which uses the modified \texttt{GADGET-3} code. We also include to the comparison the Overwhelmingly Large Simulations \citep[OWLS,][]{Schaye2010} based on the \texttt{GADGET-3} code and modified by \citet{Rahmati2014}. All the models may not be representative of a full sample of simulated galaxies. With this comparison, we intend to explore alternatives for GEAR and trace how different prescriptions affect galaxy properties.

To help in the comparison of those different models, we briefly highlight the main 
properties and difference of their physical model. 
AURIGA and TNG50 use baryon particles with an initial mass of 
$\sim\!4\cdot10^4M_\odot$ and $\sim\!8\cdot10^4M_\odot$ respectively.
Both models are build on top of the moving mesh \texttt{AREPO} code and share very similar
physics. The interstellar medium is described by the subgrid
model \citep{springel2003}, in which star-forming gas is treated as a two phase medium. 
They also both inject stellar feedback as energy-driven winds with however slightly different 
parameters.
The EAGLE simulations is sightly less resolved ($\sim\!2\cdot 10^6M_\odot$) as such,
it marginally resolve the Jeans scales in the warm ISM.
The stellar feedback is based on a stochastic thermal feedback approach
\citep{shaye2012} that efficiently trigger outflows without the need to turn off 
radiative cooling temporarily.

With a stellar mass resolution of $10^3\,\rm{M_\odot}$, GEAR has the highest resolution. 
GEAR is also the only code that resolve explicitly the multiphase nature of cold gas with 
temperatures down to $10\,\rm{K}$, including hydrogen self-shielding. 

\section{Methods}
\label{sec:methodology}

\begin{figure*}[t!]
\includegraphics[width=1.0\linewidth]{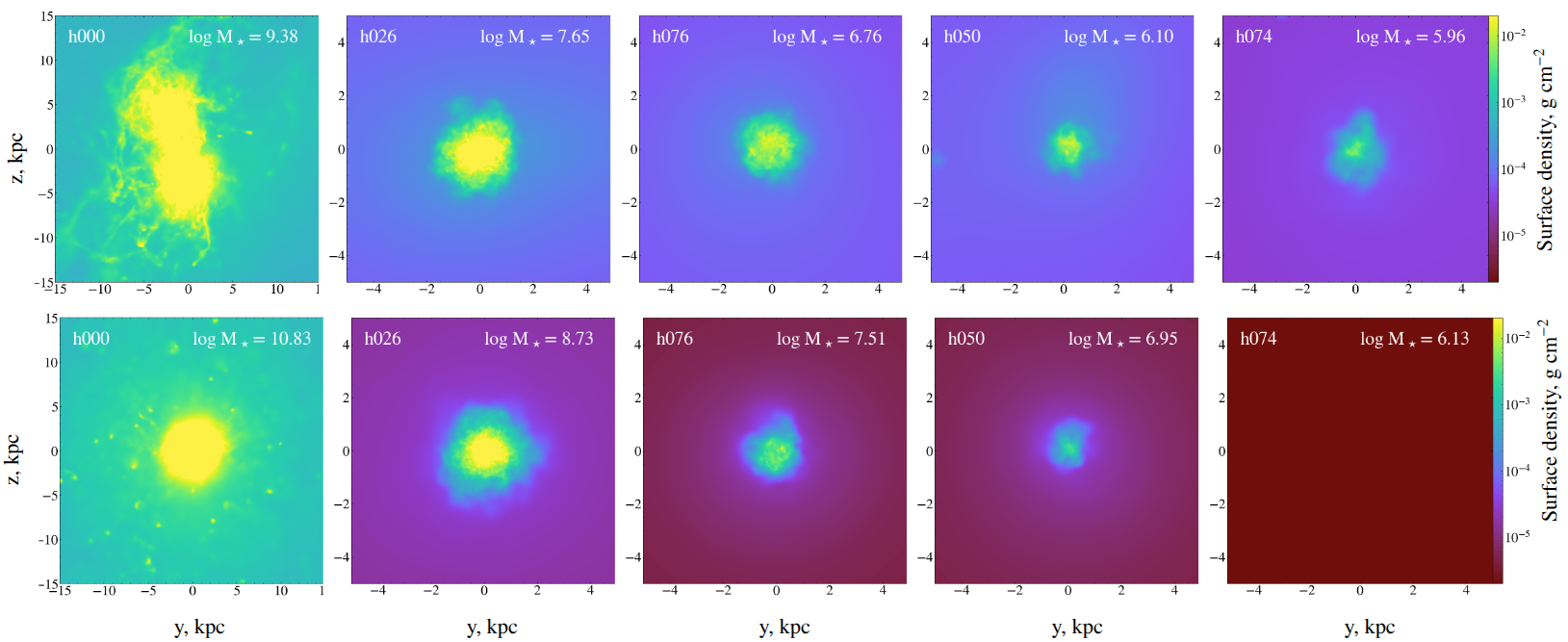}
\caption{The surface density of all gas for the five GEAR galaxies \protect\citep[][]{Revaz2018, Roca-Fabrega2021} at redshifts 4.12 (upper panel) and 0.33 (lower panel). For h074, the lower rightmost panel is empty, as all the gas has been removed by photoevaporation.}
    \label{fig:trident_imgs_z4_z0}
\end{figure*}

First, for each simulated galaxy, we determine a characteristic radius within which most of the gas belonging to a given galaxy is enclosed.
From Fig. \ref{fig:trident_imgs_z4_z0} showing the surface density of gas in the simulated galaxies, one can see that gas particles are mostly concentrated within $\sim$2 kpc from the center in dwarf galaxies and $\sim 5-10$ kpc in the MW-like galaxy h000.

The galaxy h000 has an elongated shape at $z=4.12$ (upper leftmost panel in Fig. \ref{fig:trident_imgs_z4_z0}), because at that moment it was made of two merging galaxies that doubled the volume occupied by the gas. The galaxies have merged completely by $z=$4.
For the least massive galaxy h074 (lower rightmost panel in Fig. \ref{fig:trident_imgs_z4_z0}), 
due to the presence of the UV background, the ISM is gradually fully ionized and heated, reaching a point where its specific energy exceeds the gravitational binding energy of the dwarfs. Consequently, it gradually leaves the system (evaporates) and the gas properties at z=1.15 and 0.33 cannot be studied.

\subsection{Determining H I column densities}
\label{sec:column_densities}

To obtain H I column densities in the simulated galaxies, we
use TRIDENT, a parallel Python-based open-source code for producing synthetic observations from astrophysical hydrodynamical simulation outputs \citep[][]{Hummels2017}. By creating a ray of light along a chosen line of sight crossing a galaxy, TRIDENT generates an absorption spectrum
arising from interaction with intervening material that makes up the ISM. An illustration of the method is given in Fig. \ref{fig:DLA_scheme}. To build statistics, we create light rays crossing each simulated galaxy at different galactocentric distances with a step of 0.1 kpc for dwarf galaxies or 0.2 kpc for the galaxy h000. We repeat this procedure for six different orientations spaced by 30 degree around x, y, and z axes, in such a way mimicking 18 viewing angles. Then, at each galactocentric distance we calculate the H I column density averaged over the 18 lines of sight. After all, we define the DLA size of each model galaxy as the volume within which the H I column density is high enough to correspond to DLA (i.e. log N(H I)$>20.3$).

\subsection{Determining velocity widths}
\label{sec:velocity_widths}

To determine velocity widths of the TRIDENT made absorption lines, \dvn, in the simulated galaxies, we apply the method described in \citet{Ledoux2006}. Because the GEAR simulations do not provide the number densities for ionized species of the Fe II, S II, Si II, and O I ions (normally used to obtain \dvn\, from DLA observations), we created them from the total metallicity using the TRIDENT function \texttt{add\_ion\_fields}, assuming %Fields are added assuming 
collisional ionization equilibrium and photoionization in the optically thin limit from a redshift-dependent metagalactic UV background. To compute each ion fraction as a function of density, temperature, metallicity, and redshift, TRIDENT uses the CLOUDY software \citep[see for the details][]{Hummels2017}.

Then along each line of sight we create absorption spectra produced by transitions from the ground states of the ions Fe II, S II, Si II, and O I (according to the list in Table \ref{tab:ions_list}). Among the lines, we select those that are neither strongly saturated nor too weak, and transform them %. We select lines that meet these requirements and transform their wavelengths 
into the velocity space. To make it more realistic and comparable to the observed data \citep[][]{Velichko2024}, when creating the spectrum, we choose the resolution $\Delta v$ to be 6 km s$^{-1}$, which corresponds to $R \simeq 50000$ and is comparable to UVES as well as add Gaussian random noise and specify the signal-to-noise (S/N) ratio to be equal to 30. Then we calculate the apparent optical depth
and determine $\Delta v_{90}$ as $v(95\%) - v(5\%)$, as shown in Fig. \ref{fig:dv90_determination}, where $v(5\%)$ and $v(95\%)$ are the relative velocities corresponding to the five percent and 95 percent percentiles of the cumulative apparent optical depth.

\subsection{Determining [Fe/H] and [Mg/Fe]}
\label{sec:fe_h-mg_fe}

To analyze chemical properties of gas in simulated galaxies we use the parallelized Python package pNbody \citep[][]{Revaz2013m}\footnote{\url{https://obswww.unige.ch/~revaz/pNbody/}}, a toolbox to manipulate and display interactively very large N-body systems. 
The GEAR output
includes abundances of some chemical elements produced by stars and released into the ISM through Type Ia and II supernova explosions.
To compare with the results obtained from observations by \citet{Velichko2024}, we are mainly interested in analyzing Fe and $\alpha$-elements.
To obtain the element abundances contained in gas in the simulated galaxies, we select gas particles located inside the DLA volume defined from the distribution of H I column densities, as described in Sect. \ref{sec:column_densities}, and compute the mean and dispersion of the relative abundances [Fe/H] and [Mg/Fe] \citep[where Mg is used as a proxy for $\alpha$-elements, ][]{Revaz2018}.

\section{Results and discussion}
\label{sec:results}

\subsection{Chemical properties}
\label{sec:chemical_properties}

\subsubsection{[$\alpha$/Fe] vs. [M/H]}
\label{sec:fe_h-mg_fe_discussion}

In our previous work, \citet{Velichko2024}, we have analyzed abundance patterns in DLA galaxies, separated our golden sample (see the data description in Sect. \ref{sec:data_alpha_Fe-M_H}) into two groups based on the absorption kinematics, one with the velocity widths \dvn\,$< 100$ \kms\, (low \dvn) and the other with \dvn\,$> 100$ \kms\, (high \dvn). We used \dvn\, as a proxy for the galaxy mass due to lack of measurements of the latter (see Sect. \ref{sec:vel-met_rel}). To quantify the breakpoint in \dvn, \citet{Velichko2024} used the empirical relation obtained by \citet{Arabsalmani2016}, according to which \dvn\, of 100 \kms\, corresponds to the galaxy stellar mass of 10$^9 M_\odot$\footnote{We test this relation in Sect. \ref{sec:results_dv90-mass}}. 
In \citet{Velichko2024} we found a quantitative difference in the distribution of [$\alpha$/Fe]\nucl\, vs. [M/H]\tot\, between low- and high-\dvn\, subsamples such that less massive galaxies show an $\alpha$-element knee at lower metallicities than more massive galaxies.

\begin{figure}
\centering
   \includegraphics[width=1.0\linewidth]{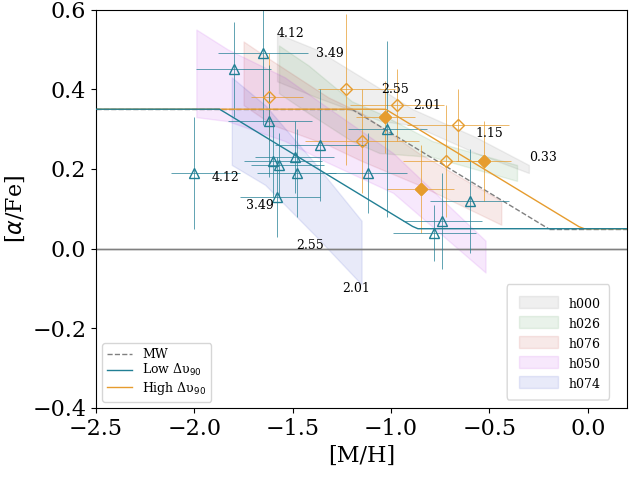}
  \caption{[$\alpha$/Fe] vs. metallicity. Shaded area: [Mg/Fe] versus [Fe/H] in gas of five simulated galaxies h000, h026, h076, h050 and h074 at $z$ from 4.12 to 0.33 (see the numbers). In the case of h074, there are no gas particles left at $z<2.01$. Orange open diamonds and blue open triangles are the values [$\alpha$/Fe]\nucl\, obtained by \protect\citet{Velichko2024} from the observations of the ISM in QSO-DLAs in the high- and low-\dvn\, subsamples, respectively. Filled symbols mark DLAs masses of which have been determined by \protect\citet{Moller2020} or \protect\citet{Christensen2014} from observations in emission of the DLA counterparts (see Table \ref{tab:DLA_measured}). The blue and orange curves are three-piecewise fits to the data for the low- and high-\dvn\, subsamples obtained in \protect\citet{Velichko2024}. The gray dashed curve shows the $\alpha$-element enhancements obtained for the MW by \protect\citet{McWilliam1997}. }
\label{fig:Mg_Fe-Fe_H}
\end{figure}

Figure \ref{fig:Mg_Fe-Fe_H} shows the [$\alpha$/Fe]\nucl\, vs. [M/H]\tot\, relations of the GEAR simulated galaxies (as 1$\sigma$ confidence intervals), together with the DLA values that we measured in \citet{Velichko2024}. We use this comparison to constrain the masses of the DLA galaxies.
The simulated data show a trend in metallicity with mass: as the galaxy mass increases, the area tend to shift toward higher metallicity because more massive galaxies enrich their surrounding gas with metals more rapidly, which is consistent with theoretical expectations and observations. Indeed, in \citet{Velichko2024}, by dividing the QSO-DLA sample into low- and high-\dvn\, subsamples, we have found that more massive galaxies showed the high-$\alpha$ knee to be at higher metallicity compared to less massive galaxies. Overall, the observed abundance ratios almost completely overlap with the output of the simulations, so we conclude that the $\alpha$-element abundances observed in our sample of DLA galaxies are consistent with stellar masses in the ranges of $10^6 - 10^8$ M$_\odot$ (h074, h050, h076) and $10^8 - 10^{11}$  M$_\odot$ (h076, h026, h000) for the low- and high-\dvn\, subsamples, respectively. 

\subsubsection{[$\alpha$/Fe] vs. specific star formation rate}

\begin{figure*}
\centering
   \includegraphics[width=0.7\linewidth]{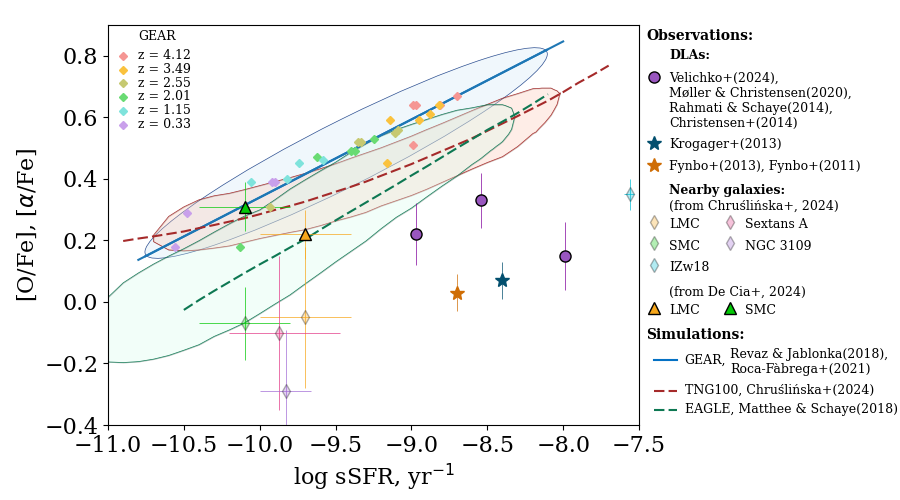}
  \caption{ [$\alpha$/Fe] or [O/Fe] vs. specific star formation rate in gas. Small filled diamonds represent [O/Fe] from GEAR \citep[][]{Revaz2018,Roca-Fabrega2021} which we fitted with the blue solid line
  and a 3$\sigma$ probability distribution function (blue shaded area). For comparison, the brown and green dashed curves, with corresponding contours, show [O/Fe] vs. sSFR from the TNG100 \citep[][]{Chruslinska2024} and EAGLE \citep[][]{Matthee2018, Chruslinska2024} cosmological simulations, respectively.
  The measurements of [$\alpha$/Fe] from absorption spectra of DLAs are taken from \citet{Velichko2024} (circles) and from \citet{Krogager2013, Fynbo2011, Fynbo2013}, corrected for dust depletion by us (stars). The measurements of sSFRs have been obtained by \citet{Moller2020, Christensen2014, Rahmati2014, Krogager2013, Fynbo2013, Fynbo2011}. The data for nearby galaxies (LMC, SMC, IZw 18, Sextans A, NGC 3109) are taken from \protect\citet{DeCia2024} and/or \protect\citet{Chruslinska2024}. All the abundances are converted to the \protect\citet{Asplund2021} solar scale.}
\label{fig:Mg_Fe-sSFR}
\end{figure*}

The specific star formation rate (sSFR), i.e. the ratio between the production rate of stars and the stellar mass accumulated over time ${\rm sSFR} = \frac{\rm SFR}{M_\star}$ (see Fig. \ref{fig:time-sSFR}), is expected to correlate with [$\alpha$/Fe] \citep[][]{Chruslinska2024}, as we see in Fig. \ref{fig:Mg_Fe-sSFR}. Theoretically, this is explained by the fact that, on the one hand, the SFR sets the amount of stars that explode as CCSNe and produce more $\alpha$-elements, on the other hand, cumulative stellar mass $M_\star$ is responsible for the continuous production of delayed SNe Ia enriching the ISM with more iron. Therefore, SFR/$M_\star$ characterize the ratio between the rate of CCSNe and SNe Ia. The [O/Fe] vs. sSFR correlation was confirmed based on the EAGLE \citep[][ green dashed curve in Fig. \ref{fig:Mg_Fe-sSFR}]{Matthee2018} and TNG100 \citep[][red dashed curve in Fig. \ref{fig:Mg_Fe-sSFR}]{Chruslinska2024} cosmological simulations.
We also obtained the relation within GEAR (small diamonds, blue solid line, and 3$\sigma$ blue confidence area in Fig. \ref{fig:Mg_Fe-sSFR}).

The GEAR, TNG100 and EAGLE simulations show slightly different behavior of [O/Fe] vs. sSFR. In Fig. \ref{fig:Mg_Fe-sSFR}, galaxies evolve from the upper right to the bottom left corner. Early on, galaxies actively form stars, while their stellar masses are still quite low, and the $\alpha$-element abundances are quite high because of the prevalence of CCSNe. With time, sSFR decreases, together with [$\alpha$/Fe] caused by increasing contribution of SNe Ia.

Observed data for five DLA galaxies are shown by stars and circles in Fig. \ref{fig:Mg_Fe-sSFR}.  The measurements of [$\alpha$/Fe] obtained from absorption spectra of QSO-DLA sources and corrected for dust depletion are taken from \citet{Velichko2024}. We also use the element abundances measured by \citet{Fynbo2013, Krogager2013, Fynbo2011} from DLA spectra to calculate [$\alpha$/Fe] corrected for dust depletion by applying the method from \citet{Velichko2024}. The measurements of stellar masses and SFRs for the corresponding DLA counterparts are taken from \citet{Moller2020, Christensen2014, Rahmati2014, Krogager2013, Fynbo2013, Fynbo2011}. The data points for six nearby galaxies (LMC, SMC, IZw 18, Sextans A, NGC 3109) shown by faint lozenges are taken from \citet{Chruslinska2024} where the values of [O/Fe] are obtained from emission-line diagnostics and the values of sSFRs calculated using the data from \citet{Skibba2012} for SMC and LMC; \citet{Hunter2010} and/or \citet{Woo2008} for Sextans A and NGC3109; \citet{Zhou2021} for IZw 18. We also include the values of [$\alpha$/Fe] obtained by \citet{DeCia2024} from observations of gas in absorption in SMC and LMC (triangles). 

Most of the observed points are systematically lower than the relations predicted by the simulations. The factors behind these discrepancies in [$\alpha$/Fe] likely differ according to the measurement specifics. (i) DLAs most often trace the outskirts of galaxies on scales of several tens of kiloparsecs and do not necessarily probe regions of recent star formation. Therefore, [$\alpha$/Fe] observed in DLAs can be systematically lower than expected in simulated star-forming galaxies. (ii) The O and Fe measurements in nearby galaxies reported by \citet{Chruslinska2024} are based on different indicators: oxygen abundances are obtained using the direct gas-phase H II region–based method while iron abundances in most cases (except for IZw 18) are determined from the spectra of young, bright stars such as blue or A-type supergiants. 
The emission-line diagnostics applied for this purpose can result in significant random and systematic uncertainties, so that variations among different calibrations up to 0.7 dex were found \citep[][]{Kewley2008}. Despite this, the values of [O/Fe] obtained by \citet{Chruslinska2024} are broadly consistent with the EAGLE simulations given the large uncertainties.

\citet{DeCia2024} employ a fundamentally different approach to studying the chemical composition of the ISM in SMC and LMC which is based on observations of the warm neutral medium in absorption. Using the ``relative'' method, the authors obtained the values of [$\alpha$/Fe] corrected for dust depletion, which are higher than the [O/Fe] measurements from \citet{Chruslinska2024} by 0.52 and 0.41 dex for SMC and LMC, respectively. The data reported by \citet{DeCia2024} is consistent with the simulations shown in Fig. \ref{fig:Mg_Fe-sSFR}. \footnote{We note that, although our focus is on issues related to abundance determinations, uncertainties also exist in the measurements of SFR and $M_\star$. These include, among other factors, systematic effects arising from the choice of SFR tracer, the assumed IMF, applied dust corrections, and the stellar population synthesis model used in certain stellar mass estimation methods \citep[][]{Chruslinska2024}.}

Furthermore, the shape of the [O/Fe] vs. sSFR relation is determined by a number of uncertain factors. \citet{Chruslinska2024} found that discrepancies between simulations are mainly attributed to the differences in the assumed SN Ia delay-time distribution and the relative SN Ia and CCSN formation efficiency and metal yields (which are the most difficult to set for the metal-poor CCSNe progenitors). The feedback model and other processes, such as merging, gas recycling, do not have a major effect on the average [O/Fe]-sSFR relation \citep[][]{Chruslinska2024}. Constraining this relation with observational data is quite challenging because of large systematic errors (when using emission-line measurements) or incomparable observations (DLAs). The gamma-ray burst (GRB) afterglows are expected to be more suitable to constrain the simulations than DLAs, because the former probe inner part of galaxies, but there are rarely enough abundance measurements to get [$\alpha$/Fe] reliably \citet{Bolmer2019}.

\subsection{H I gas distribution}
\label{sec:gas_dist}

\begin{figure}
\centering
 \includegraphics[width=1.0\linewidth]{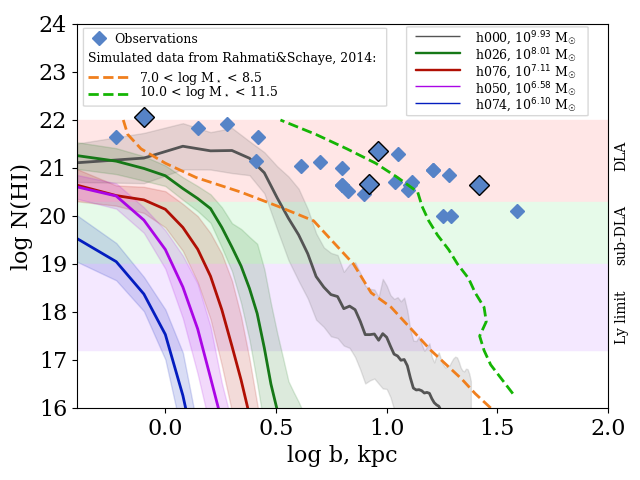}
 \caption{H I column densities vs. impact parameter at $z>1.9$. Solid curves and 1$\sigma$ confidence area correspond to the \textsc{GEAR} simulations \protect\citep[][]{Revaz2018, Roca-Fabrega2021} at $z=2.55$ averaged over 18 viewing angles. Blue diamonds are high-$z$ observations from \protect\citet{Christensen2014, Moller2020} and compilations of observed data by \protect\citet{Rahmati2014, Krogager2020}. The highlighted symbols show the objects belonging to the golden sample from \protect\citet{Velichko2024}. Dashed curves are the data from hydrodynamical simulations performed with the use of the GADGET-3 code by \protect\citet{Rahmati2014} in the stellar mass ranges of $7.0<$ log$(M_\star/M_\odot)<8.5$ (orange) and $10.0<$ log$(M_\star/M_\odot)<11.5$ (green).
 The horizontal colored stripes indicate intervals of the H I column density corresponding to DLA, sub-DLA and Ly-limit systems (see text for details).}
\label{fig:logHI-b}
\end{figure}

We applied the procedure described in Sect. \ref{sec:column_densities} to all selected snapshots of the simulated galaxies, i.e. at redshifts 4.12, 3.49, 2.55, 2.01, 1.15, 0.33.
Fig. \ref{fig:logHI-b} shows column densities of H I averaged over 18 viewing angles versus galactocentric distances in the GEAR simulated galaxies taken at redshift 2.55. The shaded areas correspond to 1$\sigma$ confidence intervals. The horizontal colored stripes in the background indicate intervals of the H I column density corresponding to DLA \citep[20.3 cm$^{-2}\leq$ log $N$(H I) $<$ 22 cm$^{-2}$,][]{Wolfe1986}, sub-DLA (19 cm$^{-2}\leq$ log $N$(H I) $<$ 20.3 cm$^{-2}$)\footnote{According to the nomenclature of \protect\citet{Prochaska2004}, these systems are called ``super Lyman limit system'' (SLLS).} and Lyman-limit systems (17.2 cm$^{-2}\leq$ log $N$(H I) $<$ 19 cm$^{-2}$)\footnote{The systems with log $N$(H I) $<$ 17.2 cm$^{-2}$ refer to as Lyman-$\alpha$ forest absorbers}. %\textbf{Those HI column densities are compared} with the observations of DLAs from \citet{Velichko2024}.

Masses and/or impact parameters of four galaxies, corresponding to DLAs, which we studied in \citet{Velichko2024}, Q0528$-$250b, Q2206$-$199a, Q2243$-$605, and Q1135$-$0010, have been measured in the works of \citet{Christensen2014} or \citet{Moller2020}. The data are summarized in Table \ref{tab:DLA_measured} and highlighted in Fig. \ref{fig:logHI-b}.

We see in Fig. \ref{fig:logHI-b} that the DLA size increases with the galaxy mass. As expected, the most massive GEAR galaxy h000 contains more gas, so its DLA size ($\sim\!4$ kpc) is higher compared to dwarf galaxies ($<2$ kpc). Note that it has less gas in its center compared to less massive galaxies due to a merging event it underwent at $z=2.55$, so the gas distribution was perturbed. For the least massive galaxy h074 with the stellar mass of $10^{6.1}M_\odot$, even in the central part the gas is not dense enough for the galaxy to appear as a DLA. Thus, we establish the lower stellar mass limit for the galaxies responsible for DLAs to be $\sim10^{6.5}M_\odot$ at $z=2.55$. Compared to observations of high-$z$ ($z>1.9$) DLA counterparts from \citet{Christensen2014, Rahmati2014, Moller2020, Krogager2020} shown by blue diamonds, the gas in the \textsc{GEAR} simulations is not extended enough to reproduce the measured values. \citet{Strawn2024} also analyze the galaxy h000 using Trident. However, they focus on the highly-ionized gas in the circumgalactic medium, and therefore these results are not directly comparable to ours. In Fig. \ref{fig:logHI-b}, we also show the data obtained from the hydrodynamical simulations performed by \citet{Rahmati2014} with the use of the GADGET-3 code in the stellar mass ranges of $7.0<$ log$(M_\star/M_\odot)<8.5$ and $10.0<$ log$(M_\star/M_\odot)<11.5$ shown by the orange and green dashed curves, respectively.
\citet{Rahmati2014} use a different approach to searching for simulated DLA counterparts, according to which the HI absorbers associated with surrounding galaxies are linked to the main galaxy.

There is an overall agreement between the \citet{Rahmati2014} simulated DLAs and the observations. This may suggest that DLA with large impact parameters may be associated also with surrounding galaxies (and they may have multiple counterparts). On the other hand, \citet{Rahmati2014} show also that, when restricting to galaxies with the range of masses and star formation of detected DLA counterparts, they find similar impact parameters. Another important aspect to consider is that observed DLA counterparts are biased against small galaxies, which are typically too close in projection to the background QSO to be detected.

\begin{figure}
\centering
 \includegraphics[width=1.0\linewidth]{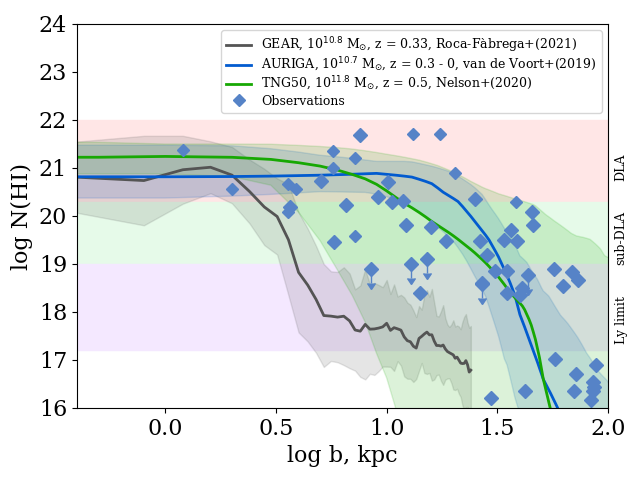}
 \caption{H I column densities vs. impact parameter at $z\sim0$. Gray curve: the \textsc{GEAR} galaxy h000 \citep[log $M_\star = 10.8 M_\odot$][]{Roca-Fabrega2021} taken at $z = 0.33$. Blue curve: the median 2D radial profile of log N(H I) for an AURIGA galaxy at $z=0-0.3$ with stellar mass of log$M_\star = 10.7 M_\odot$ resimulated by \citet{vandeVoort2019}. Green curve: cumulative covering fractions of H I from the TNG50 simulations of a galaxy with log$M_\star = 11.8 M_\odot$ \citep[][]{Nelson2020}. The low-$z$ observations taken from \protect\citet{Christensen2014, Moller2020, Kulkarni2022, Weng2023, Berg2023} are shown by blue diamonds. }
\label{fig:logHI-b_comparison}
\end{figure}

In Fig. \ref{fig:logHI-b_comparison}, we compare the H I column density distribution for h000 at $z=0.33$ with those from the AURIGA and TNG50 hydrodynamical simulations at $z$ about zero.
The blue curve shows the median 2D radial profile of log N(H I) for AURIGA galaxies at $z=0-0.3$ with stellar mass of log$M_\star = 10.7 M_\odot$ obtained by \citet{vandeVoort2019}. The stellar mass of the AURIGA galaxy is comparable to that of h000, but the gas is more extended, leading to its DLA size to be 10 times larger: $\sim$30 kpc (AURIGA) versus $\sim3$ kpc (h000). Note that the central H I column densities in these two galaxies coincide. The green curve shows the cumulative covering fractions of H I from the post-processed TNG50 simulations of a galaxy with log$M_\star = 11.8 M_\odot$ at $z=0.5$ \citep[][]{Nelson2020}. It is an order of magnitude more massive than the AURIGA galaxy, but their DLA sizes coincide within their confidence intervals. 

The values measured from observations in emission of DLA counterparts by \citet{Christensen2014}, \citet{Kulkarni2022} and \citet{Moller2020} are also provided in Fig. \ref{fig:logHI-b_comparison}.
Most sources have been detected at impact parameters higher than predicted even for the most massive \textsc{GEAR} galaxy h000. 
Unlike the TNG/Auriga models, the \textsc{GEAR} galaxy h000 has a highly concentrated center, which was also shown by \citet{Strawn2024}. 
Some of the reasons for this discrepancy include: 
(i) the stellar feedback.
Compared to TNG/Auriga, as well as EAGLE, GEAR has lower feedback coupling efficiency.
The TNG/Auriga models explicitely generate outflows during stellar feedback events 
by forcing particles to exit the galaxy without hydrodynamical interaction
\citep{grand2017,pillepich2018}. Similar outflows are obtained in EAGLE with the
stochastic thermal feedback approach \citep{shaye2012}.
These winds helps in the removal of low angular momentum material, promoting the 
formation of more extended systems.
On the contrary, GEAR employs thermal supernova feedback coupled with a delayed cooling
phase \citep{Stinson2006}. While this scheme demonstrated to be sufficient to reproduce
the stellar mass halo mass relation for classical dwarfs \citep{Revaz2018}, it does not
lead to extended more massive systems.
(ii) The ISM model.
Unlike \textsc{GEAR}, the AURIGA and TNG50 models do not include the cooling of cold gas, which is artificially kept at a medium temperature of $\sim$10$^4$K.

Primarily, the AURIGA and TNG50 simulations reproduce the observed distribution of the absorbers with impact parameter more accurately because of accounting for satellite galaxies. This is consistent with the findings reported in \citet{Weng2024} for low-$z$ objects that the central galaxy is responsible only for absorbers at the smallest impact parameters $b<0.5R_{\rm vir}$.

\renewcommand{\arraystretch}{1.2}
\begin{table*}[]
    \centering
    \caption{Properties of the DLAs from the sample used by \citet{Velichko2024}.} 
    \begin{tabular}{lcccccccc}
\hline
QSO            & $z_{\rm abs}$ & log N(H I)    & $b$         & log $M_\star$& SFR$^{\rm [R14]}$ & \dvn  & [M/H]\tot$^{\rm [V24]}$       & [$\alpha$/Fe]$^{\rm [V24]}$\\
           &               & [cm$^{-2}$]    & [kpc]       & [$M_\odot$] & [$M_\odot$yr$^{-1}$] & [\kms]  &   [dex]     & [dex]  \\
\hline
Q0528$-$250b   &  2.811        & 21.35$\pm$0.07 &  9.2$\pm$0.2$^{\rm [C14]}$   &   9.22$^{\rm [C14]}$  &     17        &   304        &  -0.85$\pm$0.17 & 0.15$\pm$0.11 \\
Q2206$-$199a   &  1.921        & 20.67$\pm$0.05 &  8.4$^{\rm [M20]}$   &   9.45$\pm$0.30$^{\rm [M20]}$ &      3        &   136        &  -0.53$\pm$0.14 & 0.22$\pm$0.10 \\
Q2243$-$605   &  2.331        & 20.65$\pm$0.05 &  26.0$^{\rm [M20]}$   &   10.1$\pm$0.1$^{\rm [M20]}$ &        36      &   173        &  -1.03$\pm$0.15 & 0.33$\pm$0.09 \\
Q1135$-$0010   &  2.207       & 22.05$\pm$0.10 &  0.8$^{\rm [M20]}$   &    -           & 25   &   168        &  -0.97$\pm$0.24 & 0.36$\pm$0.09 \\
\hline
    \end{tabular}
    \tablefoot{The masses and/or impact parameters have been determined by \tablefoottext{\textnormal{[C14]}} - \protect\citet{Christensen2014} or \tablefoottext{\textnormal{[M20]}} - \protect\citet{Moller2020}. Total metallicities and [$\alpha$/Fe] are taken from \tablefoottext{\textnormal{[V24]}} - \protect\citet{Velichko2024}. SFRs are from the compilation by \tablefoottext{\textnormal{[R14]}}-\citet{Rahmati2014}.}
    \label{tab:DLA_measured}
\end{table*}

\subsection{Velocities, masses, and metallicities}
\label{sec:velocities}

Figure \ref{fig:dv90_fromspectra} shows the velocity width versus impact parameter in the five simulated galaxies \citep[][]{Revaz2018, Roca-Fabrega2021} $z=$2.55. It increases with the galaxy mass, which is expected because, as mentioned in Sect. \ref{sec:vel-met_rel}, \dvn\, is a proxy for the galaxy dynamic mass. It also varies within the galaxies, gradually decreasing with impact parameter. We computed the mean and dispersion of velocity widths within DLA volumes of the simulated galaxies at each $z$, that allows us to trace the changes of \dvn\, on the stellar mass and metallicity of gas, as well as investigate the mass -- metallicity relation.

Compared to observations, velocity widths obtained within the simulated galaxies seem to be underestimated, especially in outer parts of the galaxies, possibly due to insufficiently strong stellar feedback in the \textsc{GEAR} code, as mentioned in Sect. \ref{sec:gas_dist}. 
On the other hand, the highest \dvn,
measured in some DLAs can be overestimated because of significant contributions from components other than the rotational motion expected for their stellar masses from the stellar Tully-Fisher relation (sTFR) obtained in the work of
\citet{Ubler2017}
\footnote{According to the data from \citet{Ubler2017}, sTFR is log$M_\star=3.56~{\rm log}V_{\rm circ} + 2.25$} 
which is shown in Fig. \ref{fig:dv90_mass} by a gray line \citep[see also the discussion in ][]{Arabsalmani2018}. For example, in case of the DLA counterpart detected by \citet{Christensen2014} (the green diamond highlighted in Fig. \ref{fig:dv90_fromspectra}), according to sTFR, its velocity width should be only 30\% of the measured value if it were only due to rotation.

The velocity spread obtained from gas observations in local galaxies, LMC \citep[$\sim\!100$ \kms,][]{Poudel2025} and SMC \citep[$\sim\!90$ \kms,][]{Murray2019}, is roughly consistent with \dvn\, predicted by the GEAR simulations.

\subsubsection{$\Delta v_{90}$ vs. stellar mass}
\label{sec:results_dv90-mass}

\begin{figure}
\centering
   \includegraphics[width=1.0\linewidth]{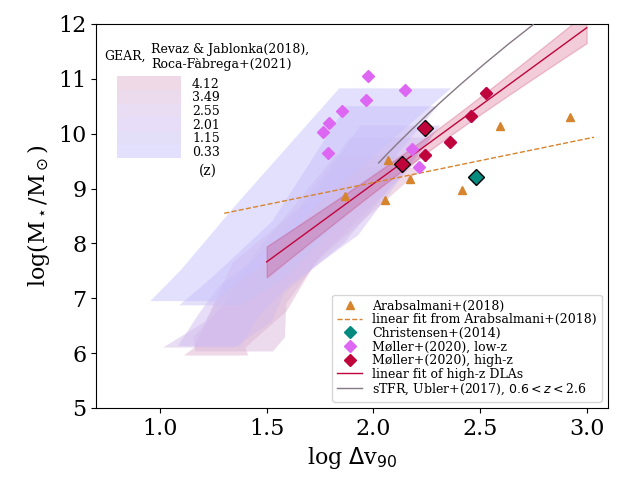}
  \caption{$\Delta v_{90}$ vs. stellar mass. Filled areas correspond to the simulated data from \protect\citet{Revaz2018} and \protect\citet{Roca-Fabrega2021}. Orange triangles are observations of GRBs used by \protect\citet{Arabsalmani2018} to obtain the relation (orange dashed line). The observations of DLAs from \protect\citet{Moller2020} are separated into low-$z$ ($\leq0.1$, magenta diamonds) and high-$z$ ($>1.9$, red diamonds) galaxies. Turquoise diamond is observations of \protect\citep{Christensen2014}. The highlighted symbols correspond to the sources in the golden sample from \protect\citet{Velichko2024}. Red and blue lines with 95\% confidence intervals are the linear fitting over high-$z$ DLAs and high-$z$ DLAs $+$ GRBs subsamples, respectively.}
\label{fig:dv90_mass}
\end{figure}

In our previous work \citep[][]{Velichko2024}, we used the $\Delta v_{90}$ vs. stellar mass relation obtained by \citet{Arabsalmani2018} based on observations of seven gamma-ray burst (GRB)-selected galaxies (orange triangles and orange dashed fitting line in Fig. \ref{fig:dv90_mass}), and estimated that \dvn\, equal to 100 \kms\, (the breakpoint between low- and high- mass galaxies) corresponds to $\sim 10^9M_\odot$. In this work, we test this relation by comparing it with the simulated data and other observations, as shown in Fig. \ref{fig:dv90_mass}. The fitting line obtained by \citet{Arabsalmani2018} is inconsistent with both the simulations and observations of DLAs from \citet{Moller2020} and disk galaxies \citep[][]{Ubler2017}. \citet{Arabsalmani2018} assumed that apparently galaxies with the highest \dvn\, have significant contributions from components other than the rotational motion (as we discussed above).
This leads to the fitting line obtained in their work being too shallow.

According to the data from \citet{Moller2020}, the high-$z$ DLAs ($z>1.9$, red diamonds) tend to be systematically shifted toward higher \dvn\, with respect to the low-$z$ DLAs ($z\leq1.9$, magenta diamonds) which may indicate an evolution of the \dvn\, vs. $M_\star$ relation with redshift. This is consistent with the \textsc{GEAR} simulated data
and can be explain by the fact that at higher redshifts galaxies appear to be more turbulent. The fraction of interacting objects may also be higher.
Compared to the orange trend from \citet{Arabsalmani2018}, fitting the high-$z$ \citet{Moller2020} subsample (red curve) gives a slope which is more consistent with both the sTFR obtained by \citet{Ubler2017} for massive star-forming disk galaxies and the \textsc{GEAR} simulated data. 

\subsubsection{$\Delta v_{90}$ vs. metallicity}

\begin{figure}
\centering
   \includegraphics[width=1.0\linewidth]{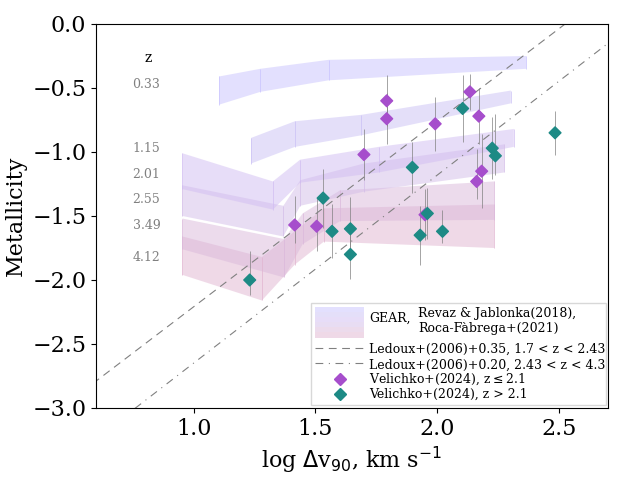}
  \caption{Velocity widths versus metallicity relation. Shaded area: the simulated data from \protect\citet{Revaz2018, Roca-Fabrega2021} at six redshifts from $z=$4.12 to 0.33 (numbers in gray) with vertical lines marking different models. Symbols show the data obtained from observations: total metallicities [M/H]\tot\, corrected for dust depletion by \protect\citet{Velichko2024} and \dvn\, found in the literature at redshift ranges $z\leq2.1$ (violet diamonds) and $z>2$ (green diamonds). Gray lines show the best fit of data on 70 DLAs at redshifts $1.7<z<2.43$ (dashed line) and $2.43<z<4.3$ (dash-dotted line) from \protect\citet{Ledoux2006}, biased in metallicity by$+0.2$ and $+0.35$ dex for the high- and low-$z$ subsamples, respectively, to roughly take into account the dust depletion.}
\label{fig:dv90-M_H}
\end{figure}

The \dvn\,-- metallicity relation (see Fig. \ref{fig:dv90-M_H}) was discovered in the work of \citet{Ledoux2006} based on observations of 70 DLAs located at redshifts in the range $1.7<z<4.3$, and it is the consequence of an underlying mass--metallicity relation. The authors found a strong redshift evolution of the relation for the $1.7<z<2.43$ and $2.43<z<4.3$ subsamples (dashed and dash-dotted lines in Fig. \ref{fig:dv90-M_H}, respectively). Due to the fact that metallicities [X/H] in \citet{Ledoux2006} were traced by volatile metals without taking into account dust depletion, we apply an offset in metallicity of $+0.2$ and $+0.35$ dex for the high- and low-$z$ subsamples to roughly correct the data, based on the mean adjustments found in \citet{DeCia2016}.
The shaded areas show the relations obtained from the gas component in the \textsc{GEAR} simulated galaxies \citep[][]{Revaz2018, Roca-Fabrega2021} at six redshifts (marked by gray numbers). The diamonds correspond to the observations of DLAs at $z\leq2.1$ (violet diamonds) and $z>2.1$ (green diamonds) from \citet{Velichko2024} where the metallicities have been corrected for dust depletion, [M/H]\tot, and velocity widths are taken from the literature.

There are some inconsistencies between the observations and simulations. First, at the low-\dvn\, end the simulated galaxies reach quite high metallicities which are not observed in real galaxies. The reasons for this could be that (i) these low mass systems are simply missed in observation because the gas is not very extended, so, the probability to have a quasar in a line of sight crossing the gas is low, especially at low $z$; (ii) the stellar feedback of the \textsc{GEAR} model is not strong enough, leading to slightly too metal rich systems. On the contrary, at the high-\dvn\, end model metallicities are not high enough to reproduce the slope that we expect from the observations \citep[shown by the gray dashed line obtained by][]{Ledoux2006}. It is likely that \textsc{GEAR} may not be ideally calibrated in the MW-mass regime.
Metallicity of gas in the galaxy h074 (with the lowest stellar mass and \dvn) is higher than in the more massive galaxy h050. This is possible due to a slightly higher sSFR at the moment of formation of h074 compared to h050.

\subsubsection{The mass -- metallicity relation}

The mass (or luminosity)--metallicity relation (MZR) is known since the work of \citet{Lequeux1979}. One possible explanation for its existence is that outflows, generated by starburst winds, blow metal-enriched gas out of low-mass galaxies more efficiently due to their shallower potential well, which slows down metal enrichment \citep[][]{Tremonti2004, DeLucia2004, Finlator20081}. Alternatively, the ``galaxy downsizing'' scenario has been proposed \citep[][]{Juneau2005, Feulner2005, Franceschini2006}, according to which massive galaxies form most of their stars rapidly and at high redshifts while low-mass galaxies evolve slower. Alternatively, \citet{Koppen2007} proposed a different mechanism for the origin of this dependence, which is based on a variation of the integrated galactic initial mass function on the SFR, which, in turn, varies in galaxies of different masses.

\begin{figure}
\centering
   \includegraphics[width=1.0\linewidth]{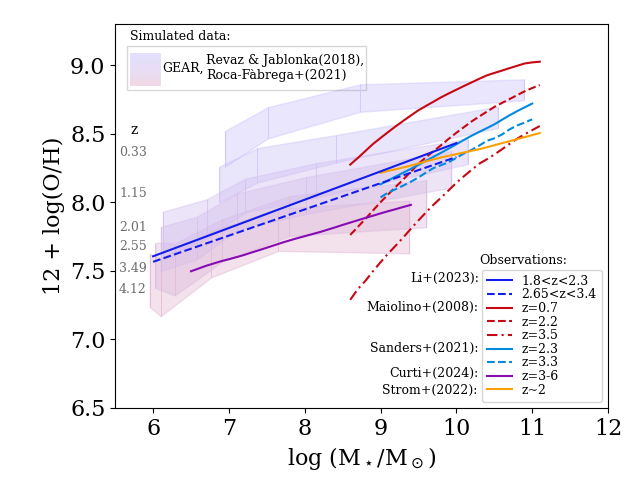}
  \caption {The mass -- metallicity relation using 12$+$log(O/H) as the metallicity tracer. Shaded bands: simulated data with 1$\sigma$ confidence intervals \protect\citep[][]{Revaz2018, Roca-Fabrega2021} at redshifts marked by the numbers in gray. The curves present the observed MZR obtained in the literature.
  Dark blue: \protect\citet{Li2023} from the study of dwarf galaxies
  at $1.8<z<2.3$ and $2.65<z<3.4$ with JWST/NIRISS. 
  Red: \protect\citet{Maiolino2008} by reprocessing the data from \protect\citet{Tremonti2004,Kewley2008,Savaglio2005, Erb2006} at $0.1<z<2.2$; as well as from the VLT near-IR observations performed with SINFONI at redshifts $3<z<5$.
  Blue: \protect\citet{Sanders2021} from the MOSDEF survey of galaxies at $z\sim$2.3 and 3.3. 
  Purple: \protect\citet{Curti2024} from the analysis of low-mass galaxies at $3<z<6$ observed with JWST/NIRSpec. Orange: \protect\citet{Strom2022} from a sample of 195 star-forming galaxies at $z\sim2$ from the Keck Baryonic Structure Survey.
  }
\label{fig:mass-M_H}
\end{figure}

MZRs obtained from observations are mainly based on the gas-phase metallicity which is traced by the oxygen abundance, and usually quoted as 12$+$log(O/H). 
Fig. \ref{fig:mass-M_H} compares the MZRs, with 12$+$log(O/H) being used as the metallicity tracer, obtained from the \textsc{GEAR} simulations at $0.33\leq z \leq 4.12$ (shaded bands of 1$\sigma$ confidence intervals) with those obtained from observations of galaxies at different redshifts and mass ranges (colored curves). Here, we use the model abundance ratio O/H contained in the gas particles falling within DLA sizes of the GEAR galaxies.

\citet{Maiolino2008} have found a strong evolution of MZR with redshift (at $0.1<z<5$) in the galaxy mass range $8.5<M_\star/M_\odot<11$ by collecting and reprocessing data from \citet{Tremonti2004} and \citet{Kewley2008} at $z\sim0.1$; from \citet{Savaglio2005} at $0.4<z<1$; from \citet{Erb2006} at $z\sim2.2$; as well as from the new near-IR observations of galaxies at $3<z<5$ performed at VLT with SINFONI (see red curves in Fig. \ref{fig:mass-M_H}). 
\citet{Sanders2021} investigated the MZRs using representative samples of $\sim300$ galaxies at $z\sim2.3$ and $\sim150$ galaxies at $z\sim3.3$ from the MOSDEF survey \citep[][]{Kriek2015}, a program that used the Multi-Object Spectrometer For Infrared Exploration \citep[MOSFIRE,][]{McLean2012}. From these data, the evolution with redshift is less prominent than it was obtained by \citet{Maiolino2008}, as well as the slope is shallower (blue curves in Fig. \ref{fig:mass-M_H}). \citet{Strom2022} obtained an even shallower slope from a sample of 195 star-forming galaxies at $z\sim2$ from the Keck Baryonic Structure Survey (orange curve in Fig. \ref{fig:mass-M_H}).
In the low-mass regime ($M_\star/M_\odot<9.5$), the MZRs were obtained in the works of \citet{Li2023}, who used observations of 51 dwarf galaxies at $z=2-3$ by JWST/NIRISS imaging and slitless grism spectroscopic observations (dark blue curves in Fig. \ref{fig:mass-M_H}), as well as \citet{Curti2024} based on observations of galaxies at $3<z<6$ with JWST/NIRSpec (purple curve in Fig. \ref{fig:mass-M_H}). 

\begin{figure}
\centering
   \includegraphics[width=1.0\linewidth]{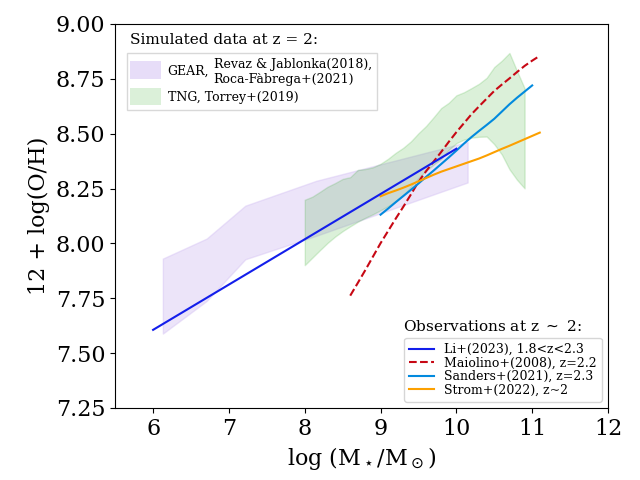}
  \caption {The mass -- metallicity relation at $z\sim2$. Blue shaded area: simulated data from \protect\citet{Revaz2018, Roca-Fabrega2021} at $z=2.01$. Green shaded area: the MZR obtained by \protect\citet{Torrey2019} from the TNG simulations at $z=2$. The colored curves present the observed MZR obtained in the literature at $z\sim2$ (the same line styles as in Fig. \ref{fig:mass-M_H}). 
  }
\label{fig:mass-M_H_z2}
\end{figure}

Overall, observations of distant galaxies show some variations in the behavior of MZR and its dependence on redshift. The evolution with redshift obtained in the works of \citet{Li2023} and \citet{Sanders2021} is smaller compared to the \textsc{GEAR} simulations. To simplify the comparison, from the data displayed in Fig. \ref{fig:mass-M_H}, we select only those corresponding to $z\sim2$ (see Fig. \ref{fig:mass-M_H_z2}). We add the MZR obtained by \citet{Torrey2019} from the TNG100 simulations at $z=2$ (the shaded green band of the $1\sigma$ confidence interval in Fig. \ref{fig:mass-M_H_z2}). Although the extent of the gas distribution might be underestimated in \textsc{GEAR} (see the discussion in Sect. \ref{sec:gas_dist}), the chemical modeling is robust and also consistent with the TNG100 simulations. \textsc{GEAR} is particularly important, because it constrains the MZR in the low-mass range, which is not targeted by such projects as Illustris-TNG and AURIGA. Despite some differences in slopes, the MZRs obtained from observations by \citet{Li2023, Sanders2021} and \citet{Strom2022} are in agreement with the simulations within $1\sigma$ confidence intervals.  The slope obtained by \citet{Maiolino2008} is steeper than both simulations and other observations. 

%*******************************************************************************
\section{Conclusions}
\label{sec:conclusions}

In this work, we conducted a comprehensive comparison of chemical and dynamical properties of DLA galaxies studied in \citet{Velichko2024, Weng2023, Berg2023, Kulkarni2022, Moller2020, Christensen2014, Rahmati2014, Krogager2013, Fynbo2013} and \citet{Fynbo2011} with the chemodynamical \textsc{GEAR} simulations performed by \citet{Revaz2018} and \citet{Roca-Fabrega2021} that cover the stellar mass range $6.1\leq M_\star/M_\odot\leq10.8$ and redshifts from 4.12 to 0.33. We additionally incorporate chemodynamical properties of model galaxies obtained from the Illustris-TNG \citep[][]{Nelson2020}, EAGLE \citep[][]{Matthee2018}, AURIGA \citep[][]{vandeVoort2019} and modified OWL simulations \citep[][]{Rahmati2014}.

We find that the abundance ratios [$\alpha$/Fe] and [M/H] observed in the ISM of DLA galaxies almost completely overlap with the abundance trends in gas of the simulated galaxies. From Fig. \ref{fig:Mg_Fe-Fe_H} we have concluded that the DLA galaxies in our sample from \citet{Velichko2024} may be associated with stellar mass ranges of $10^6 - 10^8$ $M_\odot$ and of $10^8 - 10^{11}$ $M_\odot$ for the low- and high-\dvn\, subsamples, respectively. The simulated galaxies also confirm that the $\alpha$-knee is at lower metallicities for lower-mass galaxies.

We confirm that high values of [$\alpha$/Fe] trace intense recent star formation in galaxies (see Fig. \ref{fig:Mg_Fe-sSFR}). 
The \textsc{GEAR}, TNG100 and EAGLE simulations show slightly different behavior of [O/Fe] vs. sSFR, but all the relations are consistent within 3$\sigma$ confidence areas. The discrepancies with the observations arise from different factors, depending on specifics such as the type of source and the abundance-determination method employed. We found good agreement between all the theoretical relations shown in Fig. \ref{fig:Mg_Fe-sSFR} and the values of [$\alpha$/Fe] determined by \citet{DeCia2024} for the LMC and SMC ISM based on observations of the warm neutral medium in absorption. The rest observations are hardly consistent with the model predictions due to the likely presence of large systematic errors up to 0.7 dex \citep[in case of observations in emission,][]{Chruslinska2024, Kewley2008} or incomparable observations \citep[DLA vs. host galaxy,][]{Rahmati2014, Weng2024}. 

We investigate the distribution of gas in the \textsc{GEAR} simulated galaxies (Fig. \ref{fig:logHI-b}) and find that they are more compact than the observed DLAs. For example, in the MW-mass galaxy h000 the high-density gas corresponding to DLA is concentrated within $\sim$4 kpc while the measured impact parameters of DLAs can be ten times larger. Considering that DLAs are commonly observed in group environments \citep[e.g,][]{Peroux2019}, one of the ways to resolve the inconsistency is to search for high density HI absorbers outside the main galaxy, which is supported in the work of \citet{Rahmati2014}. Based on the TNG50 simulations, \citet{Weng2024} have shown that at low $z$ only the absorbers observed at impact parameters smaller than 0.5$R_{\rm vir}$ are associated with the central galaxy, and this value decreases with increasing column density. Otherwise, they predominantly arise from satellite and other halo galaxies. The smaller DLA galaxies cannot be detected in emission at small impact parameters, because too close in projection to the background QSO.

The GEAR simulations exhibit a systematic redshift evolution of the \dvn\,--stellar mass relation toward lower \dvn, a trend corroborated by the observational data of \citet{Moller2020}. However, constraining the slope of the relation remains a nontrivial issue: the GEAR data reproduce well the sTFR obtained by \citet{Ubler2017},  while the slope obtained by fitting the data for high-$z$ DLAs from \citet{Moller2020} is somewhat shallower, and incorporating the measurements from \citet{Christensen2014} and particularly the values obtained by\citet{Arabsalmani2018} from GRBs, further increases the discrepancy. 
The high velocity widths \dvn\, obtained for the DLAs at high impact parameters (see also Fig. \ref{fig:dv90_fromspectra}), as well as GRBs \citep[][]{Arabsalmani2018}, can be explained by contribution from components other than rotational motion, e.g. multiple objects such as intergalactic clouds and/or satellite galaxies intersected by a DLA along an extended sightline \citep[][]{Rahmati2014, Weng2024}.

One of the main results is that the \textsc{GEAR} simulations also reproduce well the most recent observed mass--metallicity relations at $z\sim2$ (see Fig. \ref{fig:mass-M_H_z2}). 

Here we compare our results with a small set of other simulated galaxies, in the attempt of 1) validating the GEAR simulations against other simulations, in a parameter space that is more relevant for DLAs and 2) slightly extending the comparison to galaxy properties that are beyond those simulated with GEAR, to check for other potential limitations and improvements of GEAR simulations.

Overall, this is the first time when we compare in such details the chemodynamical properties of observed and simulated galaxies at $z\sim2-4$. It is quite difficult to make simulations that would perfectly reproduce all the variety of observed galaxy properties. Such a comprehensive analysis we perform in this work helps us better understand the processes governing the formation and evolution of galaxies.
The results of this paper consolidate the use of DLAs for the study of galaxy (chemical) evolution.
A full and detailed analysis and discussion of the comparison with the additional simulations, beyond GEAR, as well as the exploration of the variation of the assumptions of GEAR simulations, in particular its feedback prescription, is beyond the scope of this paper and should be addressed in future work.

\begin{acknowledgements} 
A.V., A.D.C. and J.K.K. acknowledge support by the Swiss National Science Foundation under grant 185692 funding the ``Interstellar One'' project. J.K.K. acknowledges financial support from the French Agence Nationale de la Recherche (ANR) under grant number ANR-24-CE31-7454.

We are grateful to the anonymous referee for valuable comments and suggestions, which significantly improved the manuscript.
\end{acknowledgements}

\bibliographystyle{aa} % style aa.bst
\bibliography{mybib}

\begin{appendix}

\section{Additional details of the GEAR galaxies}
\label{ap:GEAR}

\begin{figure}
\centering
   \includegraphics[width=1.0\linewidth]{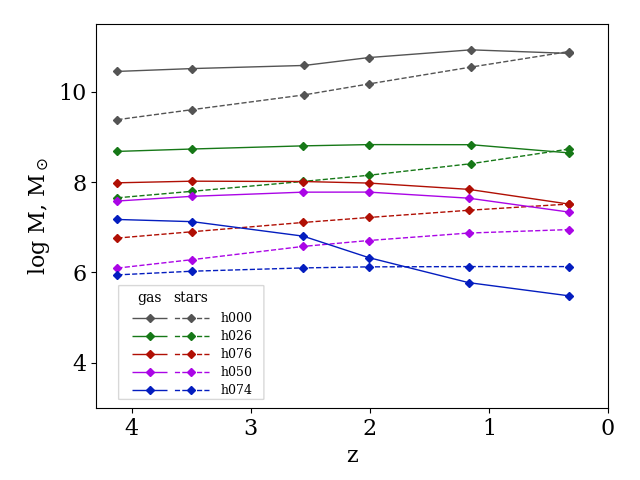}
  \caption{Redshift evolution of stellar and gas masses measured within $R_{\rm vir}$ in the five simulated galaxies. The solid and dashed curves correspond to the gas and stellar components, respectively.} 
\label{fig:logM-z}
\end{figure}

\begin{figure}
\centering
   \includegraphics[width=1.0\linewidth]{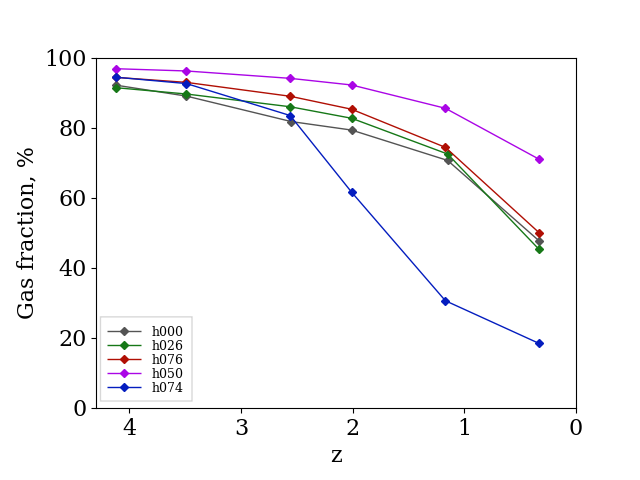}
  \caption{Fraction of all gas in the simulated galaxies $M_{\rm gas}/(M_{\rm gas} + M_\star)$. }
\label{fig:gas_fraction}
\end{figure}

Some characteristics of the GEAR galaxies are provided in Table \ref{tab:simgal_props}. Fig. \ref{fig:logM-z} shows the evolution of stellar and gas masses (measured within the virial radius $R_{\rm vir}$) of the galaxies with redshift.
The number of stars gradually increases with the galaxy evolution due to star formation processes, while the amount of gas on average gradually decreases because one part of the gas transforms into stars, another is ejected due to outflows caused by stellar feedback and UV-background heating during the reionization epoch \citep[][]{Revaz2018}. This effect is especially important for the least massive galaxy h074 with shallower potential well (blue curves in Fig. \ref{fig:logM-z}), which quickly loses almost all its gas by $z=0.33$. On the other hand, the gas reservoirs can also be replenished through gas accretion and/or merger events. For example, the gray solid curve in Fig. \ref{fig:logM-z}, corresponding to the h000 galaxy, shows a slight increase in the mass of gas between $z=2.55$ and 2.01 because of merging, which is consistent with the results shown in Fig. 2 of \citet{Strawn2024}. The gas fraction $f_{\rm gas} = M_{\rm gas}/(M_{\rm gas} + M_\star)$ gradually decreases with galaxy evolution, as shown in Fig. \ref{fig:gas_fraction}, but the four more massive galaxies remain gas-rich systems (with $f_{\rm gas} \geq 50\%$) down to $z\sim1$, while in the h074 galaxy $f_{\rm gas}$ decreases sharply starting from $z\sim2$ and reaches 0\% by $z=0$ due to more efficient gas loss.

\renewcommand{\arraystretch}{1.0}
\begin{table*}[]
    \centering
    \caption{Properties of the GEAR galaxies at $z = 0$.}
    \begin{tabular}{ccccccc}
\hline
Model ID & log $L_V$,  & log $M_\star$,& log $M_{\rm gas}$,& [Fe/H]$_{\rm gas}$, & [Mg/Fe]$_{\rm gas}$, & Ref.\\
         & [$L_\odot$] & [$M_\odot$]   & [$M_\odot$]       & [dex]               &  [dex]              & \\
\hline
h000     &  8.65      &  10.8          &   10.3            &    $-0.30\pm0.05$   &   $0.20\pm0.01$     & 2 \\
h026     &  8.65      &  8.73          &   8.66            &    $-0.36\pm0.08$   &   $0.19\pm0.02$     & 1 \\
%h019     &  8.46      &  8.64          &   8.45            &                     &                     & 1 \\
h076     &  7.27      &  7.58          &   7.36            &    $-0.44\pm0.09$   &   $0.09\pm0.03$    & 1 \\
h050     &  6.62      &  6.98          &   7.18            &    $-0.52\pm0.11$   &   $-0.02\pm0.04$    & 1 \\
h074     &  5.70      &  6.13          &   5.59            &    $-1.15\pm0.14^*$   &   $-0.01\pm0.08^*$   & 1 \\
\hline
    \end{tabular}
    \tablefoot{The simulations are from \protect\citet{Revaz2018} (1) and \protect\citet{Roca-Fabrega2021} (2). The asterisk \tablefoottext{*} marks the values provided for $z=2.01$, because at lower redshifts there are no gas particles left within the galaxy h074.}
    \label{tab:simgal_props}
\end{table*}

\section{An illustration of a DLA}
\begin{figure}
\centering
    \begin{minipage}{0.04\linewidth}
    \raggedleft
    \vspace{-7cm} 
    \rotatebox{90}{\scriptsize z, kpc}
    \end{minipage}
\begin{minipage}[t]{0.95\linewidth}
   \includegraphics[height=7cm, trim={2cm 1.4cm 2cm 0},clip]{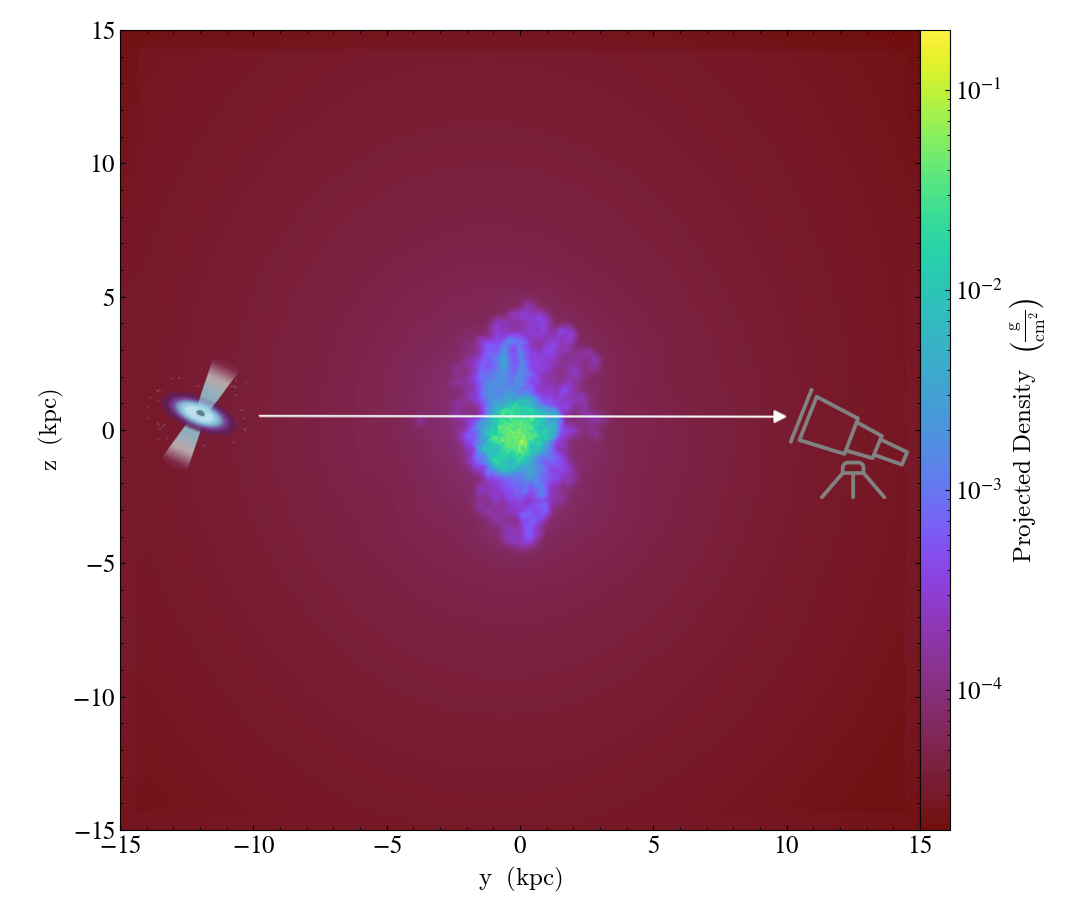}
      \put(2,70){\rotatebox{90}{\footnotesize Surface density, g cm$^{-2}$}}
\end{minipage}    
    \begin{minipage}[t]{0.95\textwidth}
    \hspace{3.8cm}{\scriptsize y, kpc }
    \end{minipage}
  \caption{Illustration of a line-of-sight passing through the simulated galaxy h026 taken at $z=2.01$. The schematic image of a quasar is taken from the \copyright\, NASA image gallery.
 }
\label{fig:DLA_scheme}
\end{figure}

\section{Determination of $\Delta v_{90}$}
\label{sec:app_lines}

\renewcommand{\arraystretch}{1.0}
\begin{table*}[]
    \centering
    \caption{List of lines used to determine $\Delta v_{90}$.}
    \begin{tabular}{lccc|lccc}
\hline
Ion & $\lambda$, [\AA] & gamma & f value        & Ion & $\lambda$, [\AA] & gamma & f value \\
\hline
Fe II & 1608.451 & 1.910000e+08 & 5.910000e-02   & S II  & 906.876  & 1.090000e+09 & 2.010000e-0  \\
Fe II & 1143.226 & 1.000000e+08 & 1.900000e-02  &  S II  & 765.684  & 9.470000e+09 & 1.250000e+00 \\         
Fe II & 1127.098 & 6.000000e+06 & 1.100000e-03  &  S II  & 764.416  & 9.550000e+09 & 8.360000e-01 \\  
Fe II & 1125.447 & 1.000000e+08 & 1.600000e-02  &  S II  & 763.656  & 9.600000e+09 & 4.190000e-01 \\        
Fe II & 1121.975 & 1.900000e+08 & 2.900000e-02  & Si II & 1808.0130 & 2.540000e+06 & 2.490000e-03  \\
Fe II & 1112.048 & 2.000000e+07 & 4.500000e-03  & Si II & 1304.3700 & 3.040000e+09 & 1.090000e+00 \\
Fe II & 1096.877 & 2.300000e+08 & 3.300000e-02  & Si II & 1260.4220 & 2.950000e+09 & 1.180000e+00 \\
Fe II & 1083.420 & 1.600000e+07 & 2.800000e-03  & Si II & 1193.2900 & 2.690000e+09 & 5.750000e-01 \\
Fe II & 1081.875 & 6.000000e+07 & 1.300000e-02  & Si II & 1190.4160 & 6.530000e+08 & 2.770000e-01 \\ 
Fe II & 1063.972 & 3.500000e+07 & 4.700000e-03  & Si II & 1020.6990 & 8.910000e+07 & 1.390000e-02 \\
Fe II & 1063.177 & 3.200000e+08 & 5.500000e-02  & Si II & 989.8730  & 6.810000e+08 & 2.000000e-01 \\ 
Fe II & 1062.153 & 1.700000e+07 & 2.900000e-03  &  O I   & 1302.168  & 3.410000e+08 & 5.200000e-02\\
Fe II & 1055.262 & 4.600000e+07 & 6.100000e-03  & O I   & 1039.230  & 9.430000e+07 & 9.160000e-03 \\
Fe II & 937.651  & 4.400000e+07 & 7.000000e-03  & O I   & 988.773   & 2.260000e+08 & 4.640000e-02 \\  
Fe II & 926.897  & 3.200000e+07 & 3.300000e-03  & O I   & 988.655   & 5.770000e+07 & 8.460000e-03 \\ 
Fe II & 926.212  & 2.600000e+08 & 3.300000e-02  & O I   & 976.448   & 3.860000e+07 & 3.310000e-03\\
S II  & 1259.519 & 5.100000e+07 & 1.820000e-02  &  O I   & 950.885   & 1.940000e+07 & 1.580000e-03  \\   
S II  & 1253.811 & 5.120000e+07 & 1.210000e-02  &   O I   & 936.629   & 1.660000e+07 & 3.060000e-03  \\ 
S II  & 1250.584 & 5.130000e+07 & 6.020000e-03  &  O I   & 929.517   & 1.060000e+07 & 1.920000e-03  \\ 
S II  & 912.736  & 1.050000e+09 & 6.530000e-02  &   O I   & 924.950   & 7.220000e+06 & 1.300000e-03 \\   
S II  & 910.485  & 1.060000e+09 & 1.320000e-01  &  &          &              &      \\    
\hline
    \end{tabular}
    \label{tab:ions_list}
\end{table*}

\begin{figure}
\centering
   \includegraphics[width=1.0\linewidth]{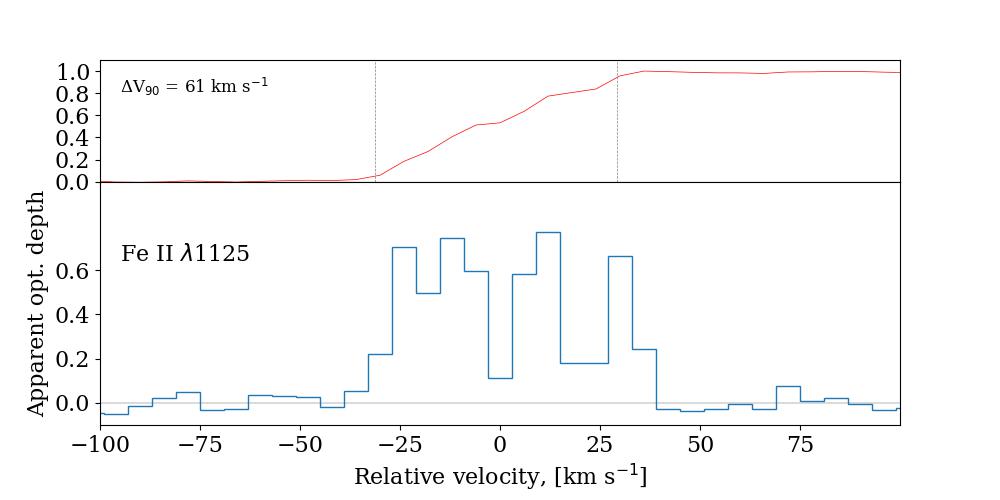}
  \caption{Measurement of the velocity width $\Delta v_{90}$ of the absorption line profile of Fe II $\lambda$1125 in the mock spectrum for the galaxy h026 taken at $z$=2.01. The spectrum is obtained along the line of sight at an impact parameter $b$ = 1.5 kpc.
  Upper panel: normalized total apparent optical depth, the dashed vertical lines show the velocity range between 5\% and 95\% of the total apparent optical depth (lower panel). The resulting $\Delta v_{90}$ is 61 km s$^{-1}$.}
\label{fig:dv90_determination}
\end{figure}

\section{Specific star formation rates in the simulated galaxies}

\begin{figure}
\centering
   \includegraphics[width=1.0\linewidth]{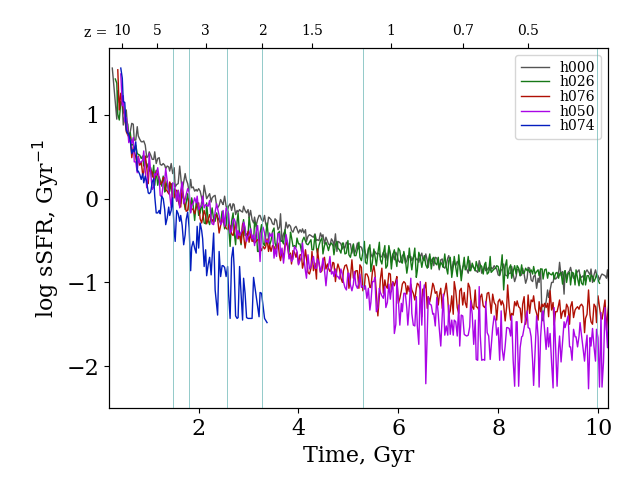}
  \caption{The sSFR vs. time for the GEAR galaxies from \protect\citet{Revaz2018} and \protect\citet{Roca-Fabrega2021}. The vertical lines mark the six redshifts $z = 4.12, 3.49, 2.55, 2.01, 1.15, 0.33$ at which we take the model galaxies.}
\label{fig:time-sSFR}
\end{figure}

\section{Velocity widths}
\begin{figure}
\centering
   \includegraphics[width=1.0\linewidth]{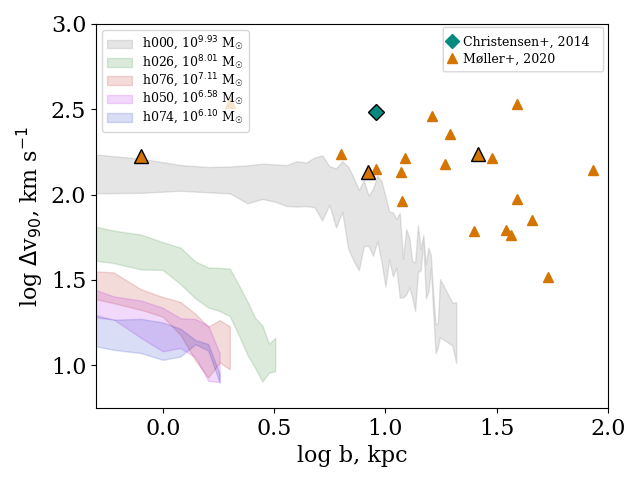}
  \caption{Velocity widths vs. impact parameter in the simulated galaxies at $z=$2.55 compared to the data obtained from observations in emission by \protect\citet{Christensen2014} (orange triangles) and \protect\citet{Moller2020} (green diamonds). The highlighted symbols correspond to the systems studied in \protect\citet{Velichko2024} (see Table \ref{tab:DLA_measured}).}
\label{fig:dv90_fromspectra}
\end{figure}

\end{appendix}

\end{document}